\documentclass[aps,prd,twocolumn,tightenlines,nofootinbib]{revtex4-1}
\usepackage[T1]{fontenc}
\usepackage{microtype}
\usepackage[textsize=scriptsize,backgroundcolor=red!70,linecolor=red]{todonotes}
\usepackage{booktabs}
\usepackage{dcolumn}
\usepackage{amsmath}
\usepackage{amsfonts}
\usepackage{amssymb}
\usepackage{graphicx}
\usepackage{hyperref}
\usepackage[all]{hypcap}
\usepackage{paralist}
\usepackage{multirow}

\usepackage{graphicx}
\usepackage{epstopdf}
\usepackage{epsfig}
\usepackage{dcolumn}
\usepackage{bm}
\usepackage{epsf}
\usepackage{dcolumn}
\def\be{\begin{equation}}
\def\ee{\end{equation}}
\def\bea{\begin{eqnarray}}
\def\eea{\end{eqnarray}}
\def\gsim{\ \rlap{\raise 2pt\hbox{$>$}}{\lower 2pt \hbox{$\sim$}}\ }
\def\lsim{\ \rlap{\raise 2pt\hbox{$<$}}{\lower 2pt \hbox{$\sim$}}\ }
\def\dslash{\kern-4pt \not{\hbox{\kern-2pt $\partial$}}}
\def\pslash{\not{\hbox{\kern-2pt p}}}

\def\l{{\rm L}}

\def \be{\beta}

\def\beq{\begin{equation}}
\def\eeq{\end{equation}}
\def\bea{\begin{eqnarray}}
\def\eea{\end{eqnarray}}
\def\ber{\begin{eqnarray*}}
\def\eer{\end{eqnarray*}}
\def\bwt{\begin{widetext}}
\def\ewt{\end{widetext}}

\def\roughly#1{\mathrel{\raise.3ex\hbox
{$#1$\kern-.75em\lower1ex\hbox{$\sim$}}}}
\def\lsim{\roughly<}
\def\gsim{\roughly>}

\def\order{\lower 1.8ex \hbox{\LARGE\~{}}}

\usepackage{bm}

\def \({\left(}
\def \){\right)}
\def \[{\left[}
\def \]{\right]}
\def \l|{\left|}
\def \r|{\right|}

\def \be{\beta}

\begin{document}
\DeclareGraphicsExtensions{.pdf,.ps}


\title{IceCube's astrophysical neutrino energy spectrum from CPT violation}


\author{Jiajun Liao}
\affiliation{Department of Physics and Astronomy, University of Hawaii at Manoa, Honolulu, HI 96822, USA}
 
\author{Danny Marfatia}
\affiliation{Department of Physics and Astronomy, University of Hawaii at Manoa, Honolulu, HI 96822, USA}

\begin{abstract}

The 6-year dataset of high-energy starting events (HESE) at IceCube indicates a spectrum of astrophysical neutrinos much softer than expected from the Fermi shock acceleration mechanism. On the other hand, IceCube's up-going muon neutrino dataset and Fermi-LAT's gamma-ray spectrum point to an $E^{-2}$ neutrino spectrum. If the HESE data above 200 TeV are fit with the latter flux, an excess at lower energies ensues, which then suggests a multicomponent spectrum.
We show that the HESE dataset can be explained by a single $E^{-2}$ power-law neutrino flux from a muon-damped $p\gamma$ source if neutrino interactions are modified by CPT violation. The low-energy excess is naturally explained by the pileup of events from superluminal neutrino decay, and there is no cutoff at high energies due to the contribution of subluminal antineutrinos. The best-fit scenario with CPT violation also predicts the observation of Glashow resonance events in the near future.
  
\end{abstract}
\maketitle

The discovery of high-energy astrophysical neutrinos in the IceCube detector at the South Pole has opened a new window into astrophysics and neutrino physics~\cite{Aartsen:2013bka}. Since neutrinos are not affected by magnetic fields and have almost no interactions with matter during propagation from their source to the Earth, observation of high-energy astrophysical neutrinos has long been recognized as a powerful tool to study the origin of ultra high-energy cosmic rays; for a review see Ref.~\cite{Anchordoqui:2013dnh}. 
The latest 6-year HESE dataset is comprised of a sample of 82 events with deposited energy $E_\text{dep}$ between 30~TeV and 2~PeV, while the total expected number of atmospheric muon background events and the atmospheric neutrino background events are $25.2\pm 7.3$ and $15.6_{-3.9}^{+11.4}$, respectively~\cite{Aartsen:2017mau}. The observed neutrino events are consistent with an isotropic distribution and are found to be predominantly of extragalactic origin~\cite{Denton:2017csz}. However, a Galactic origin is still viable~\cite{Taylor:2014hya}.

\begin{figure}
\centering
\includegraphics[width=0.4\textwidth]{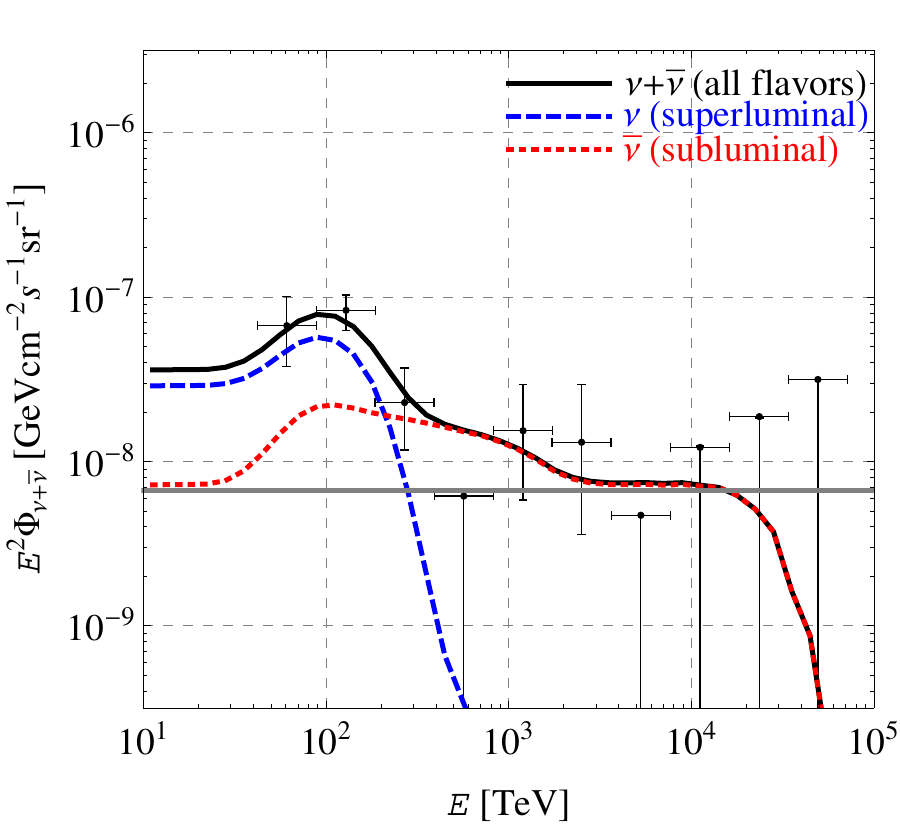}
\caption{The total flux of astrophysical neutrinos as a function of neutrino energy. The data points are extracted from 6-year HESE data assuming a $1:1:1$ flavor ratio, and an $E^{-2}$ spectrum and independent normalization in each energy bin~\cite{Aartsen:2017mau}. The gray horizontal line shows the best-fit to HESE data with deposited energy above 200~TeV for an $E^{-2}$ spectrum and $1:1:1$ flavor ratio; the total flux is given by $E^2\Phi=0.66\times 10^{-8} \text{GeV}\text{cm}^{-2}\text{s}^{-1}\text{sr}^{-1}$. The black solid curve corresponds to the total flux on Earth for the best fit CPTV scenario. The blue dashed (red dotted) curve corresponds to the superluminal neutrino (subluminal antineutrino) component. Here we assume a muon-damped $p\gamma$ source with an $E^{-2}$ spectrum. The initial fraction of the superluminal neutrinos is 80\% and $E_{th}=1.2$ PeV. The final total flux is normalized to $E^2\Phi=3.6\times 10^{-8} \text{GeV}\text{cm}^{-2}\text{s}^{-1}\text{sr}^{-1}$ at $E= 10\text{ TeV}$.
}
\label{fig:flux}
\end{figure}

The extracted flux in the energy range $60\text{ TeV}<E<50\text{ PeV}$ is shown in Fig~\ref{fig:flux}. A likelihood fit of the data with a single unbroken power-law flux, $E^{-\gamma}$, gives a spectral index $\gamma=2.92^{+0.33}_{-0.29}$, 
which is in conflict with the Fermi-LAT gamma-ray flux~\cite{Ackermann:2014usa} that requires $\gamma\lesssim 2.1-2.2$ for neutrinos produced via hadronic collisions~\cite{Murase:2013rfa}. Also, the analysis of 8-year IceCube data of up-going muon neutrinos with
 $E_\nu\gtrsim 120$~TeV yields a harder spectrum with $\gamma=2.19\pm 0.10$~\cite{Aartsen:2017mau}, which is consistent with the HESE flux above 200~TeV. 
An $E^{-2}$ spectrum is generally expected from the standard Fermi shock acceleration mechanism~\cite{Gaisser:1990vg}. However, fitting the HESE data above 200~TeV with an $E^{-2}$ spectrum with a fixed normalization results in an excess between 40~TeV and 200~TeV, as can be seen in Fig.~\ref{fig:flux}. The maximum local statistical significance of the excess is 2.6$\sigma$~\cite{Chianese:2017nwe}.  Imposing a prior from the up-going muon neutrino flux for the HESE data at high energies, a nonzero softer component is then preferred~\cite{Aartsen:2017mau}. In fact, an analysis using only shower events in the 4-year HESE data shows $3\sigma$ evidence for a break in the astrophysical neutrino spectrum~\cite{Anchordoqui:2016ewn}. 

It has been speculated that a multicomponent flux arising from new physics may be responsible for features in the IceCube spectrum~\cite{He:2013zpa}.
Not surprisingly, since the events detected at IceCube have the highest neutrino energies observed,  they provide a unique opportunity to probe fundamental physics in the neutrino sector that cannot be reproduced in the laboratory. 
In this Letter, we assume a single unbroken $E^{-2}$ source spectrum and demonstrate how CPT violation (CPTV) reproduces the high energy neutrino spectrum observed by IceCube.

{\bf CPT and Lorentz invariance violation.}
Violations of CPT and Lorentz invariance originating at the Planck scale are well motivated in quantum gravity theories~\cite{Kostelecky:1988zi}, and can be described in the effective field theory framework of the standard model extension~\cite{Colladay:1998fq}. Since rotational symmetry is highly constrained experimentally, we only consider the consequences of operators that preserve it. Specifically, we focus on modifications of the kinematics of particle interactions. For simplicity, we assume Lorentz invarince violation (LIV) and CPTV only occur in the neutrino sector, and that the new interactions are flavor blind so as to avoid the stringent constraints from neutrino oscillation experiments. Before proceeding, we note that CPTV implies LIV, but not vice versa~\cite{greenberg}.

\begin{figure}
\centering
\includegraphics[width=0.45\textwidth]{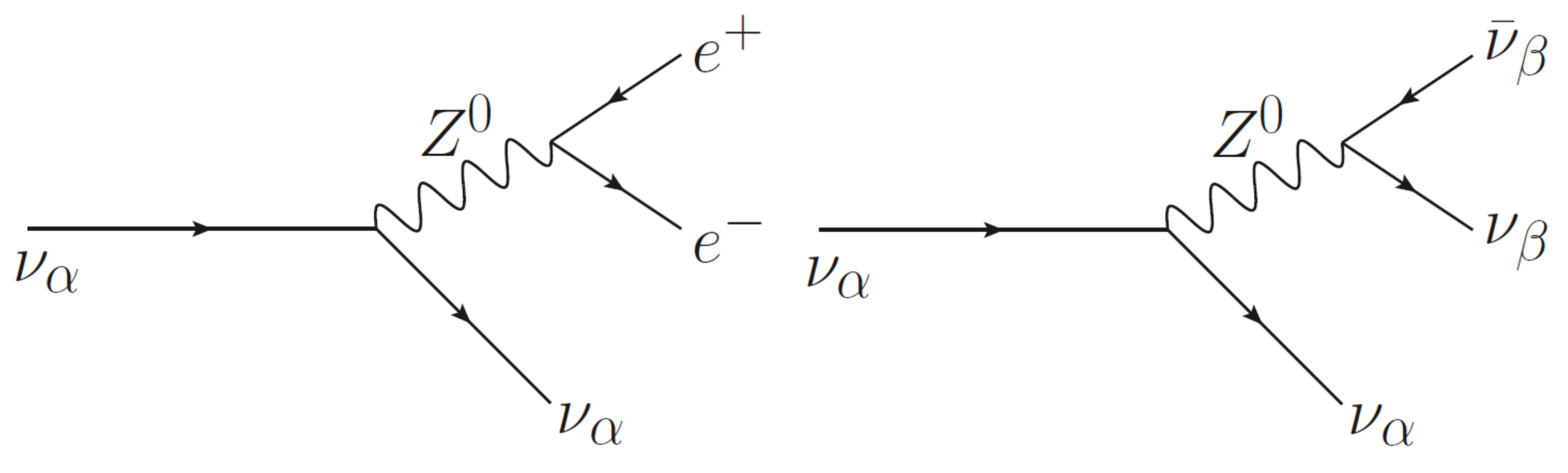}
\caption{The Feynman diagrams for vacuum electron-positron pair emission (left) and neutrino splitting (right) for superluminal neutrinos.
}
\label{fig:feyndiag}
\end{figure}

LIV leads to a modified dispersion relation for neutrinos, and consequently, some reactions that are forbidden kinematically become allowed if neutrinos are superluminal~\cite{Cohen:2011hx}. For high energy superluminal neutrinos, the dominant reactions (shown in Fig.~\ref{fig:feyndiag}) include vacuum electron-position pair emission (VPE), $\nu \to \nu + e^+ + e^-$, and neutrino splitting, $\nu_\alpha \to \nu_\alpha + \nu_\beta +\bar{\nu}_\beta$, where $\alpha,\beta=e, \mu, \tau$. We ignore the charged current contribution to VPE because the electron neutrino population on Earth is suppressed for the neutrino source we consider below.

The modified dispersion relation for neutrinos is
\begin{align}
E^2-p^2=m^2+2\delta E^2\,,
\label{eq:dispersion}
\end{align}
where the LIV parameter $\delta$ is of the form~\cite{Kostelecky:2011gq}
\begin{align}
\delta=\sum_n\kappa_n\left(\frac{E}{M_{Pl}}\right)^n\,,
\label{eq:delta}
\end{align}
which is generally suppressed by the Planck mass $M_{Pl}$. 
Terms with an even (odd) mass dimension, i.e., $n$ even or odd,  conserve (violate) CPT. In principle, the LIV effects could be the sum of all CPT-even and CPT-odd terms. Here we consider the case that a single $n$ term dominates the others. 

For the CPTV case, the dispersion relation of antineutrinos is given by Eq.~(\ref{eq:dispersion}) with $\delta \to -\delta$. Hence,
depending on the sign of $\delta$, either the neutrinos are superluminal and antineutrinos are subluminal, or vice versa~\cite{Kostelecky:2011gq}.  There is no cutoff in the neutrino spectrum for the CPTV case because of the subluminal component~\cite{Stecker:2014oxa}. Also, there is an excess in the neutrino spectrum below the redshifted threshold energy due to the pileup
 of the decay products from the superluminal component. On the other hand, for the CPT-even case, 
the neutrinos and antineutrinos are either both superluminal or both subluminal. If they are both superluminal at the source, their decay produces a cutoff at energies just above the event 
pileup~\cite{Stecker:2014oxa} which makes it impossible to explain the events above a few hundred TeV. Since the excess in the IceCube HESE data occurs below 200~TeV, we do not consider the CPT-conserving case any further.

{\bf Simulation.}
We simulate the propagation of cosmological neutrinos from the source to the Earth using Monte Carlo techniques. We assume the energy spectrum at the source follows a single power law, $E^{-2}$, and use the redshift distribution of the neutrino sources from Ref.~\cite{Stecker:2014xja}, which follows the star formation rate~\cite{Behroozi:2012iw}.  We take $\delta >0$.

For pion decay, $\pi^+ \to \mu^+\nu_\mu$, to proceed, the energy of the superluminal neutrinos produced at the source must be bounded from above.  Conservation of energy-momentum requires~\cite{Kostelecky:2011gq}
\begin{align}
2 \kappa_n\left(\frac{E}{M_{Pl}}\right)^n E^2\leq (m_\pi-m_\mu)^2\,,
\label{eq:cutoff}
\end{align} 
where $m_\pi$ and $m_\mu$ are the masses of pions and muons, respectively. Interestingly, at higher energies, the $\pi^+$ become stable and constitute a new cosmic ray primary~\cite{Anchordoqui:2014hua}. Also, VPE will only occur above a threshold neutrino energy $E_{th}$, which is given by~\cite{Stecker:2001vb}
\begin{align}
2 \kappa_n\left(\frac{E_{th}}{M_{Pl}}\right)^n E_{th}^2=4m_e^2\,,
\label{eq:Eth}
\end{align}
where $m_e$ is the electron mass. Since $\kappa_n$ depends on $E_{th}$ monotonically, we characterize the size of LIV by $E_{th}$. From Eqs.~(\ref{eq:cutoff}) and~(\ref{eq:Eth}), we see that the highest energy muon neutrino event observed imposes a lower bound on $E_{th}$, i.e.,
\begin{align}
E_{th} \geq \left(\frac{2m_e}{m_\pi-m_\mu}\right)^{\frac{2}{n+2}}E_{obs}\,,
\end{align}
where $E_{obs}$ is the observed muon neutrino energy. Note that the minimum value of the VPE threshold energy increases as $n$ increases. Since the highest energy track event observed by IceCube has a median estimated neutrino energy of 8.7 PeV~\cite{hitrack}, we obtain 
$E_{th}^{min} = 0.85$~PeV for $n=1$. This bound does not apply if the event is initiated by a subluminal antineutrino.

For a given VPE threshold energy $E_{th}$ or equivalently $\kappa_n$ (from Eq.~\ref{eq:Eth}), the upper bound on the superluminal neutrino energy in Eq.~(\ref{eq:cutoff}) can be written as
\begin{eqnarray}
E&\leq& \left(\frac{m_\pi-m_\mu}{2 m_e}\right)^{\frac{2}{n+2}}E_{th}\\
& \leq&
10.3E_{th} \ \ \   {\rm for} \  n=1\,.
\end{eqnarray}
Henceforth, we set $n=1$.
In our simulation, we assume the superluminal neutrino energy at the source is between 10 TeV and the upper bound for a given $E_{th}$. Since there is no upper bound on the energy for subluminal antineutrinos, we assume their energy lies between 10~TeV and 
100~PeV.
We then propagate the neutrinos from the source to the Earth. During propagation, superluminal neutrinos redshift and lose energy via VPE and neutrino splitting, while subluminal antineutrinos only experience redshifting.

The VPE rate depends on the choice of dynamical matrix element employed to incorporate superluminal neutrinos. We use the results of Ref.~\cite{Carmona:2012tp} for the case that the dynamical matrix element is that of special relativity but with the modified dispersion relation in Eqs.~(\ref{eq:dispersion}) and~(\ref{eq:delta}). The decay rate is~\cite{Carmona:2012tp,Stecker:2014oxa}
\begin{align}
\Gamma=\frac{G_F^2E^5}{192\pi^3}\left[(1-2s_W^2)^2+(2s_W^2)^2\right]\xi_1\kappa_1^3\frac{E^3}{M_{Pl}^{3}}\,,
\label{eq:decayrate}
\end{align}
where $s_W$ is the sine of the Weinberg angle and $\xi_1=\frac{209}{140}$~\cite{Carmona:2012tp}.
For the VPE process, the mean fractional energy loss is 0.74~\cite{Carmona:2012tp}. 

For the neutrino splitting process, we assume that the decay rate is three times that of the VPE and each of the three daughter neutrinos carries one third of the parent neutrino energy. Modifications of these assumptions have negligible effects on the shape of the neutrino spectrum at the Earth~\cite{Stecker:2014oxa}. The energy loss due to redshifting is given by
\begin{align}
\frac{\partial \log E}{\partial t} =-H_0\sqrt{\Omega_m(1+z)^3+\Omega_\Lambda}\,,
\end{align}
where the Hubble constant $H_0=67.8\text{ km }\text{s}^{-1}\text{Mpc}^{-1}$, $\Omega_\Lambda=0.7$ and $\Omega_m=0.3$~\cite{planck}. An example of the Monte Carlo result is shown in Fig.~\ref{fig:flux}.

\begin{figure}
\centering
\includegraphics[width=0.45\textwidth]{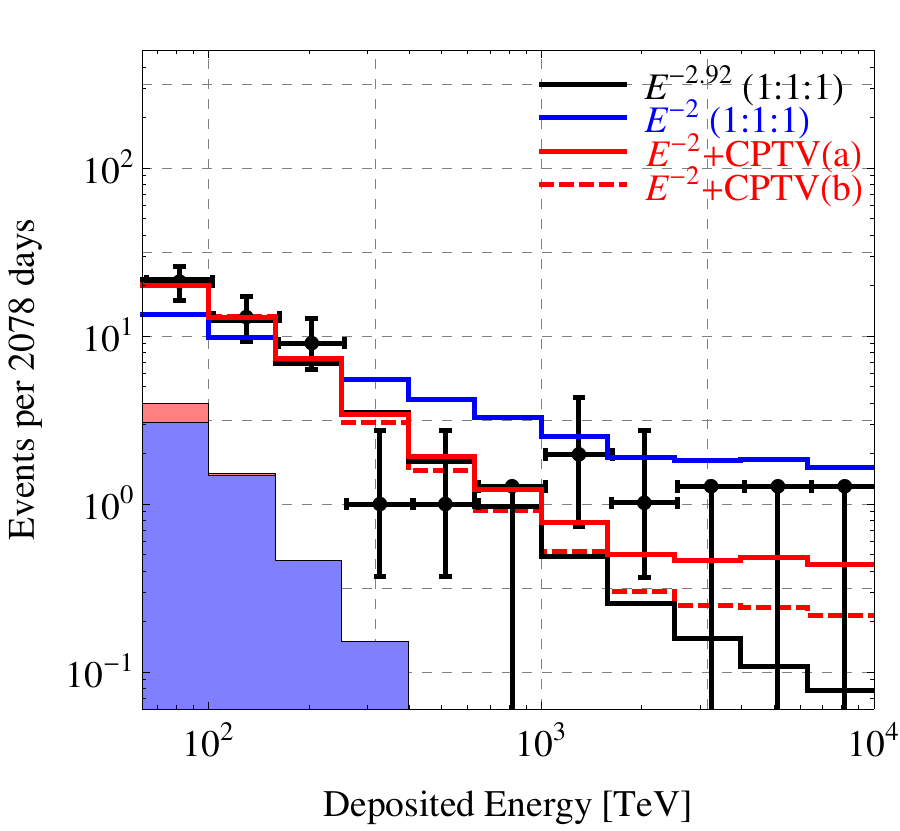}
\caption{The number of events per 2078 days as a function of the deposited energy in the IceCube detector. The data points are taken from the IceCube 6-year HESE dataset~\cite{Aartsen:2017mau}. The atmospheric muon and atmospheric neutrino backgrounds are shaded red and blue, respectively. The black (blue) curve corresponds to the best-fit scenario for a single power law of $E^{-2.92}$ ($E^{-2}$) in case (a); cases (a)/(b) exclude/include the three highest energy bins in the statistical analysis. The red solid [dashed] curve corresponds to the best-fit CPTV scenario in case (a) [(b)], with $E_{th}=1.2$ PeV and an initial fraction of superluminal neutrinos of 80\% [90\%]. For the scenarios without CPTV, a flavor ratio of $1:1:1$ is assumed on Earth. For the CPTV scenario, we assume a muon-damped $p\gamma$ source with an $E^{-2}$ spectrum.
}
\label{fig:spectrum}
\end{figure}

{\bf Analysis and results.}
We analyze the energy spectrum of the IceCube 6-year HESE dataset, which is shown in Fig.~\ref{fig:spectrum}. We extract the atmospheric muon and atmospheric neutrino background from Ref.~\cite{Aartsen:2017mau}. For the astrophysical neutrinos, we convolve the incident neutrino flux on Earth with the effective areas given in Ref.~\cite{effAreadata}, which allows us to predict the deposited energy spectrum. We use the central values for all 28 effective areas that are separated by the particle type (neutrino or antineutrino), interaction channel (charged-current deep-inelastic scattering, neutral current deep-inelastic scattering, or resonant anti-electron-neutrino/electron
scattering), and event topology (track or cascade). As a check we reproduced the IceCube prediction for the $E^{-2.92}$ flux. As in Ref.~\cite{Aartsen:2017mau}, we only consider the energy bins with deposited energy between 60~TeV and 10~PeV.

To evaluate the statistical significance of a particular scenario, we define
\begin{align}
\chi^2=&\sum_i2\left[\alpha N_i^\text{th}(r,E_{th})-(N_i^\text{data}-\beta N_i^\text{bkg})\right.  \\\nonumber
&+\left.(N_i^\text{data}-\beta N_i^\text{bkg})\ln\frac{N_i^\text{data}-\beta N_i^\text{bkg}}{\alpha N_i^\text{th}(r,E_{th})}\right]\\\nonumber
&+\sum_j2\alpha N_j^\text{th}(r,E_{th})+\left(\frac{1-\beta}{\sigma_\beta}\right)^2\,,
\end{align}
where $i=1-5,7,8$, $j=6, 9-11$, and $\alpha$ and $\beta$ are normalization parameters for the astrophysical neutrinos and the backgrounds, respectively. $N^\text{th}$ ($N^\text{data}$) [$N^\text{bkg}$] are the predicted number of astrophysical neutrino events (experimental measured number of events) [the atmospheric neutrino and muon background]. For the CPTV case, $N^\text{th}$ depends on the initial fraction of the total number of superluminal neutrinos and antineutrinos, $r$, and the VPE threshold energy $E_{th}$. Here we take $\sigma_\beta=0.26$ as a penalty for the normalization of the atmospheric neutrino and muon background, and the normalization of the astrophysical neutrinos $\alpha$ is allowed to float.

Since no events have been observed near the Glashow resonance (GR)~\cite{Glashow:1960zz} in the HESE dataset, we consider two cases: (a) the three highest energy bins are excluded from the analysis; (b) the three highest energy bins are included. We first analyze the spectrum for single power laws, $E^{-2.92}$ and $E^{-2}$, without CPTV. We assume the flavor ratio is $1:1:1$ on Earth. The $\chi^2$ of the best-fit scenarios are provided in Table~\ref{tab:chi}. We also calculate the predicted number of GR events for each best-fit scenario by using the effective areas for resonant scattering. 
The predicted spectra for the best-fit results in case (a) are shown in Fig.~\ref{fig:spectrum}. Clearly, the $E^{-2}$ spectrum is disfavored by HESE data.
%

\begin{table}[t]

\begin{center}
\begin{tabular}[c]{|l|>{\centering\arraybackslash}p{1cm}|>{\centering\arraybackslash}p{0.9cm}|>{\centering\arraybackslash}p{1cm}|>{\centering\arraybackslash}p{0.9cm}|>{\centering\arraybackslash}p{1cm}|>{\centering\arraybackslash}p{0.9cm}|}
\hline
\multirow{2}{*}{ }&\multicolumn{2}{c|}{$E^{-2.92}$$(1:1:1)$}&\multicolumn{2}{c|}{$E^{-2}$$(1:1:1)$}&\multicolumn{2}{c|}{$E^{-2}$ with CPTV} \\
\hline
Case & (a) & (b) & (a) & (b) &(a) &(b) \\
                      \hline
$\chi^2$ & 9.6 & 10.3 & 24.0 & 34.0 & 7.7 & 9.4  \\
\hline
GR events & 0.16 & 0.15 & 3.1 & 2.7 & 0.98 & 0.49 \\
\hline
\end{tabular}
\end{center}
\label{tab:chi}
\caption{The $\chi^2$ and predicted number of Glashow resonance (GR)  events for the best-fit scenarios. In case (a)/(b), the three highest energy bins are excluded from/included in the analysis. For the scenarios without CPTV, a flavor ratio of $1:1:1$ is assumed. For the CPTV scenario, we assume a muon-damped $p\gamma$ source with an $E^{-2}$ spectrum.}

\end{table}

\begin{figure}[t]
\centering
\includegraphics[width=0.4\textwidth]{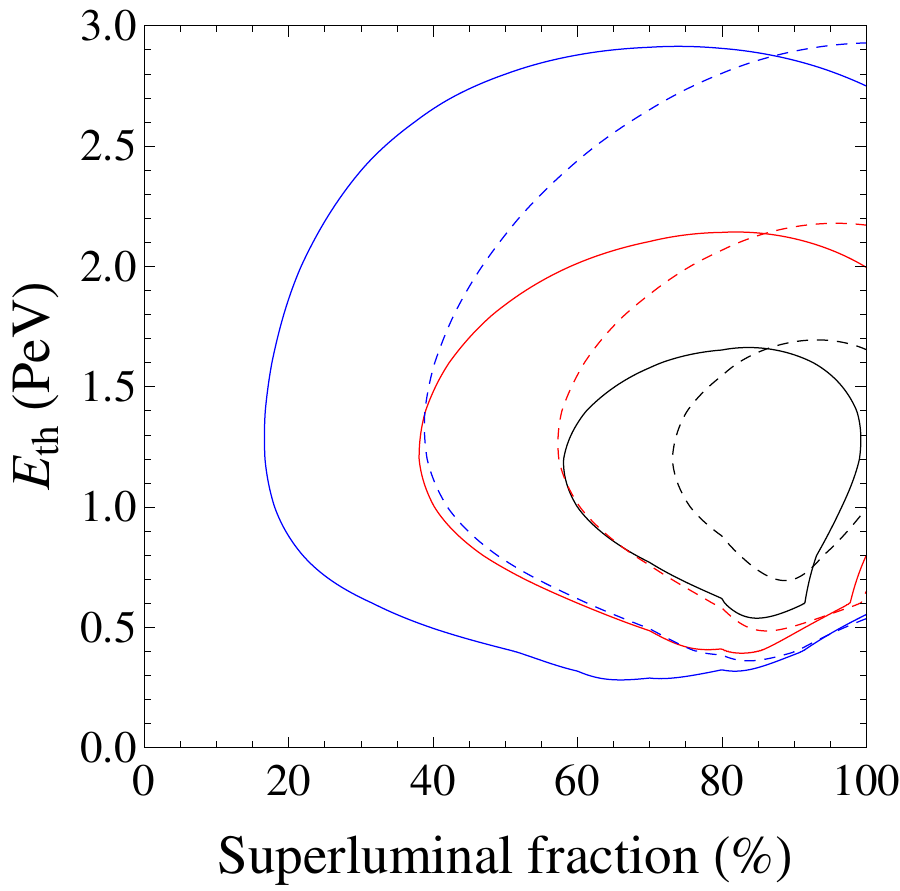}
\caption{1$\sigma$, 2$\sigma$, 3$\sigma$ allowed regions for IceCube 6-year HESE data. We assume the neutrinos are produced by a muon-damped $p\gamma$ source with an $E^{-2}$ spectrum. The solid (dashed) curves correspond to the case in which the three highest energy bins are excluded from/included in the fit. 
}
\label{fig:scan}
\end{figure}

We now assume an $E^{-2}$ spectrum at the source, and analyze the deposited energy spectrum on Earth invoking CPTV in propagation. 
We find that in order to explain the excess in the HESE dataset, a large asymmetry between the initial neutrino and antineutrino flux is required, which cannot be obtained from a $pp$ source. However, this can be achieved by considering a $p\gamma$ source with muon damping, i.e., $p\gamma\to \pi^+\to \nu_\mu$ only.  An ideal muon-damped $p\gamma$ source produces only superluminal neutrinos for $\delta>0$. However, in realistic objects like Gamma Ray Bursts and Active Galactic Nuclei, an intrinsic contamination from $\pi^-$ is expected to reduce the superluminal fraction by 20\%--33\%~\cite{Biehl:2016psj}. The contamination depends on the target photon spectrum and neutrino source model. We leave the superluminal fraction as a free parameter. The muon-damped $p\gamma$ decay chain yields a flavor composition of $\nu_e:\nu_\mu:\nu_\tau=0:1:0$ at the source and approximately $4:7:7$ at Earth due to neutrino oscillations (for both neutrinos and antineutrinos)~\cite{Barger:2014iua}.
Note that current data are not sensitive enough to discriminate between realistic flavor ratios~\cite{Aartsen:2017mau}. Hence, introducing more source modes only serves to decrease the initial fraction of superluminal neutrinos. As a corollary, we find that a $p\gamma$ 
source with partial muon damping also works.

We scan over the parameter space of the initial fraction of superluminal neutrinos $r$ and the VPE threshold energy $E_{th}$. 
The 1$\sigma$, 2$\sigma$, 3$\sigma$ allowed regions in the ($r$,$E_{th}$) parameter space are shown in Fig.~\ref{fig:scan}. 
The $\chi^2$ and the predicted number of GR events of the best-fit scenarios are listed in Table~\ref{tab:chi}. We find that the CPTV scenario with an $E^{-2}$ source spectrum gives the best fit among all three scenarios for both cases (a) and (b), and the improvement over an $E^{-2}$ spectrum without CPTV is significant.

The best-fit CPTV scenarios for case (a) and case (b) both occur at $E_{th}=1.2$ PeV, with an initial fraction of superluminal neutrinos at 80\% and 90\%, respectively. These fractions are reasonable given the expectation for $\pi^-$ contamination (which produces subluminal antineutrinos).
The predicted spectrum for the best-fit scenarios are shown in Fig.~\ref{fig:spectrum}. The best-fit CPTV scenarios provide a better fit between 300~TeV and 3~PeV compared to the $E^{-2.92}$ spectrum. 
Also, they predict a larger GR event rate than the $E^{-2.92}$ spectrum, which may become interesting since there is an indication of a 6~PeV shower event at IceCube~\cite{GS}. 

{\bf Summary.} 
We showed that the IceCube 6-year HESE dataset is well-explained by an $E^{-2}$ flux from a muon-damped $p\gamma$ neutrino source if CPT violation modifies neutrino interactions. The superluminal neutrino fraction needed to replicate the data is compatible with $\pi^-$ contamination (which only produces subluminal antineutrinos). A multicomponent flux is not needed to fit the low energy excess because superluminal neutrino decay naturally produces an event pileup below 200~TeV. Subluminal antineutrinos contribute a flux at high energies so there is no cutoff in the spectrum. The best-fit scenario with CPTV also predicts a Glashow resonance event rate at the edge of IceCube detection. 

{\it Acknowledgments.} 
We thank V.~Berezinsky, A.~Connolly and S.~Scully for helpful correspondence.
This research was supported in part by the U.S. DOE under Grant No. DE-SC0010504.

%

%




\begin{thebibliography}{99}
\bibitem{Aartsen:2013bka} 
  M.~G.~Aartsen {\it et al.} [IceCube Collaboration],
  Phys.\ Rev.\ Lett.\  {\bf 111}, 021103 (2013)
  [arXiv:1304.5356 [astro-ph.HE]];
    Science {\bf 342}, 1242856 (2013)
    [arXiv:1311.5238 [astro-ph.HE]];
      Phys.\ Rev.\ Lett.\  {\bf 113}, 101101 (2014)
      [arXiv:1405.5303 [astro-ph.HE]].


\bibitem{Anchordoqui:2013dnh} 
  L.~A.~Anchordoqui {\it et al.},
  JHEAp {\bf 1-2}, 1 (2014)
  [arXiv:1312.6587 [astro-ph.HE]].

%

\bibitem{Aartsen:2017mau} 
  M.~G.~Aartsen {\it et al.} [IceCube Collaboration],
  arXiv:1710.01191 [astro-ph.HE].

\bibitem{Denton:2017csz} 
  P.~B.~Denton, D.~Marfatia and T.~J.~Weiler,
  JCAP {\bf 1708}, no. 08, 033 (2017)
  [arXiv:1703.09721 [astro-ph.HE]].

\bibitem{Taylor:2014hya} 
A.~M.~Taylor, S.~Gabici and F.~Aharonian,
Phys.\ Rev.\ D {\bf 89}, no. 10, 103003 (2014)
[arXiv:1403.3206 [astro-ph.HE]]; 
M.~Ahlers, Y.~Bai, V.~Barger and R.~Lu,
Phys.\ Rev.\ D {\bf 93}, no. 1, 013009 (2016)
[arXiv:1505.03156 [hep-ph]];
A.~Neronov and D.~V.~Semikoz,
Astropart.\ Phys.\  {\bf 75}, 60 (2016)
[arXiv:1509.03522 [astro-ph.HE]];
A.~Palladino and F.~Vissani,
Astrophys.\ J.\  {\bf 826}, no. 2, 185 (2016)
[arXiv:1601.06678 [astro-ph.HE]];
G.~Pagliaroli, C.~Evoli and F.~L.~Villante,
JCAP {\bf 1611}, no. 11, 004 (2016)
[arXiv:1606.04489 [astro-ph.HE]];
D.~Grasso, D.~Gaggero, A.~Marinelli, M.~Taoso and A.~Urbano,
Nuovo Cim.\ C {\bf 40}, no. 3, 140 (2017).

\bibitem{Ackermann:2014usa} 
  M.~Ackermann {\it et al.} [Fermi-LAT Collaboration],
  Astrophys.\ J.\  {\bf 799}, 86 (2015)
  [arXiv:1410.3696 [astro-ph.HE]].

\bibitem{Murase:2013rfa} 
  K.~Murase, M.~Ahlers and B.~C.~Lacki,
  Phys.\ Rev.\ D {\bf 88}, no. 12, 121301 (2013)
  [arXiv:1306.3417 [astro-ph.HE]].

  
\bibitem{Gaisser:1990vg} 
  T.~K.~Gaisser,
  Cambridge, UK: Univ. Pr. (1990) 279 p

\bibitem{Chianese:2017nwe} 
  M.~Chianese, G.~Miele and S.~Morisi,
  Phys.\ Lett.\ B {\bf 773}, 591 (2017)
  [arXiv:1707.05241 [hep-ph]].

\bibitem{Anchordoqui:2016ewn} 
  L.~A.~Anchordoqui, M.~M.~Block, L.~Durand, P.~Ha, J.~F.~Soriano and T.~J.~Weiler,
  Phys.\ Rev.\ D {\bf 95}, no. 8, 083009 (2017)
  [arXiv:1611.07905 [astro-ph.HE]].

\bibitem{He:2013zpa} 
See e.g.,
  Y.~Ema, R.~Jinno and T.~Moroi,
  Phys.\ Lett.\ B {\bf 733}, 120 (2014)
  [arXiv:1312.3501 [hep-ph]];
L.~A.~Anchordoqui, V.~Barger, H.~Goldberg, X.~Huang, D.~Marfatia, L.~H.~M.~da Silva and T.~J.~Weiler,
      Phys.\ Rev.\ D {\bf 92}, no. 6, 061301 (2015)
      [arXiv:1506.08788 [hep-ph]];
        P.~S.~B.~Dev, D.~Kazanas, R.~N.~Mohapatra, V.~L.~Teplitz and Y.~Zhang,
  JCAP {\bf 1608}, no. 08, 034 (2016)
  [arXiv:1606.04517 [hep-ph]];
        D.~Borah, A.~Dasgupta, U.~K.~Dey, S.~Patra and G.~Tomar,
  JHEP {\bf 1709}, 005 (2017)
  [arXiv:1704.04138 [hep-ph]];
          A.~Bhattacharya, A.~Esmaili, S.~Palomares-Ruiz and I.~Sarcevic,
  JCAP {\bf 1707}, no. 07, 027 (2017)
  [arXiv:1706.05746 [hep-ph]];
          G.~K.~Chakravarty, N.~Khan and S.~Mohanty,
  arXiv:1707.03853 [hep-ph].
  

  
\bibitem{Kostelecky:1988zi} 
  V.~A.~Kostelecky and S.~Samuel,
  Phys.\ Rev.\ D {\bf 39}, 683 (1989);
  D.~Mattingly,
    Living Rev.\ Rel.\  {\bf 8}, 5 (2005)
    [gr-qc/0502097];
     S.~Liberati,
      Class.\ Quant.\ Grav.\  {\bf 30}, 133001 (2013)
      [arXiv:1304.5795 [gr-qc]].

\bibitem{Colladay:1998fq} 
  D.~Colladay and V.~A.~Kostelecky,
  Phys.\ Rev.\ D {\bf 58}, 116002 (1998)
  [hep-ph/9809521].
  
  \bibitem{greenberg} 
  O.~W.~Greenberg,
  Phys.\ Rev.\ Lett.\  {\bf 89}, 231602 (2002)
  [hep-ph/0201258].

\bibitem{Cohen:2011hx} 
  A.~G.~Cohen and S.~L.~Glashow,
  Phys.\ Rev.\ Lett.\  {\bf 107}, 181803 (2011)
  [arXiv:1109.6562 [hep-ph]].
  


\bibitem{Kostelecky:2011gq} 
  A.~Kostelecky and M.~Mewes,
  Phys.\ Rev.\ D {\bf 85}, 096005 (2012)
  [arXiv:1112.6395 [hep-ph]].


\bibitem{Stecker:2014oxa} 
F.~W.~Stecker, S.~T.~Scully, S.~Liberati and D.~Mattingly,
Phys.\ Rev.\ D {\bf 91}, no. 4, 045009 (2015)
[arXiv:1411.5889 [hep-ph]].  


\bibitem{Stecker:2014xja} 
  F.~W.~Stecker and S.~T.~Scully,
  Phys.\ Rev.\ D {\bf 90}, no. 4, 043012 (2014)
  [arXiv:1404.7025 [astro-ph.HE]].
  
\bibitem{Behroozi:2012iw} 
  P.~S.~Behroozi, R.~H.~Wechsler and C.~Conroy,
  Astrophys.\ J.\  {\bf 770}, 57 (2013)
  [arXiv:1207.6105 [astro-ph.CO]].
  
\bibitem{Anchordoqui:2014hua} 
L.~A.~Anchordoqui, V.~Barger, H.~Goldberg, J.~G.~Learned, D.~Marfatia, S.~Pakvasa, T.~C.~Paul and T.~J.~Weiler,
Phys.\ Lett.\ B {\bf 739}, 99 (2014)
[arXiv:1404.0622 [hep-ph]]. 
  
\bibitem{Stecker:2001vb} 
  F.~W.~Stecker and S.~L.~Glashow,
  Astropart.\ Phys.\  {\bf 16}, 97 (2001)
  [astro-ph/0102226].


\bibitem{hitrack} 
  M.~G.~Aartsen {\it et al.} [IceCube Collaboration],
  Astrophys.\ J.\  {\bf 833}, no. 1, 3 (2016)
  [arXiv:1607.08006 [astro-ph.HE]].

\bibitem{Carmona:2012tp} 
  J.~M.~Carmona, J.~L.~Cortes and D.~Mazon,
  Phys.\ Rev.\ D {\bf 85}, 113001 (2012)
  [arXiv:1203.2585 [hep-ph]].

\bibitem{planck} 
  P.~A.~R.~Ade {\it et al.} [Planck Collaboration],
  Astron.\ Astrophys.\  {\bf 594}, A13 (2016)
  [arXiv:1502.01589 [astro-ph.CO]].

\bibitem{effAreadata}
M.~G.~Aartsen et al., 2015, ``Search for contained neutrino events at energies greater than 1 TeV in 2 years of data'', http://icecube.wisc.edu/science/data/HEnu\_above1tev/

\bibitem{Glashow:1960zz} 
  S.~L.~Glashow,
  Phys.\ Rev.\  {\bf 118}, 316 (1960).


\bibitem{Biehl:2016psj} 
  S.~Hummer, M.~Ruger, F.~Spanier and W.~Winter,
  Astrophys.\ J.\  {\bf 721}, 630 (2010)
  [arXiv:1002.1310 [astro-ph.HE]].

\bibitem{Barger:2014iua} 
  V.~Barger, L.~Fu, J.~G.~Learned, D.~Marfatia, S.~Pakvasa and T.~J.~Weiler,
  Phys.\ Rev.\ D {\bf 90}, 121301 (2014)
  [arXiv:1407.3255 [astro-ph.HE]].
  
  

\bibitem{GS}
Lu Lu, TeVPA 2017, ``A new search for multi-flavour PeV neutrinos with IceCube'', https://indico.cern.ch/event/615891/contributions/2636704/.
  
\end{thebibliography}
\end{document}